\documentclass{elsart1p}
\usepackage{graphicx}
\usepackage{amssymb}
\begin{document}
\begin{frontmatter}
%
%
\title{\vspace{-0.8in}Recent Longitudinal Spin Asymmetry Measurements for Inclusive Jet Production at STAR}
\vspace{0.2in}
\author{D. Staszak, for the STAR Collaboration}
\ead{staszak@physics.ucla.edu}
\address{Dept. of Physics and Astronomy, University of California, Los Angeles, 430 Portola Plaza, Box 951547,  
Los Angeles, CA 90095-1547, USA}

\title{}
%
%
\vspace{-2.0in}
\author{}
\address{}
\begin{abstract}
We present measurements of the longitudinal spin asymmetry, $A_{LL}$, for the inclusive jet signal at STAR.  The data presented here are mid-rapidity jets in the transverse momentum range of $5<pT<35$ GeV/c and come from polarized proton-proton collisions at center of mass energies of $\sqrt{s}=200$ GeV.  We compare our measured $A_{LL}$ values to predictions derived from various parameterizations of the polarized gluon distribution function.  The results are shown to provide significant constraints for allowable 
gluon parameterizations within the measurement's kinematic Bjorken-x range of $0.03<x<0.3$.
\end{abstract}

\end{frontmatter}

One of the main objectives of the Relativistic Heavy Ion Collider (RHIC) spin program\cite{bunce} 
is to use polarized p-p collisions to determine the gluon spin contribution, $\Delta G$, to the proton's spin.  
$\Delta G$ is accessed at RHIC in a number of final state channels\cite{bunce} 
by measuring the double longitudinal spin asymmetry 
$A_{LL} = \Delta\sigma / \sigma = (\sigma^{++} - \sigma^{+-}) / (\sigma^{++} + \sigma^{+-})$,
where ++ and +- refer to the helicity states of the two proton beams.  
The focus of this contribution is on the inclusive jet channel,
which has the advantages of a relatively large cross section and little sensitivity to fragmentation functions.

First inclusive jet cross section and $A_{LL}$ results from STAR have been previously
published\cite{jetspaper} and disfavor a large positive gluon spin contribution.  
The results presented here use data taken during two extended $\sqrt{s}=200$ GeV p-p runs in 2005
and 2006 at the RHIC facility.  
The corresponding measurements of $A_{LL}$ from the 
2005\cite{2005paper} and 2006 data provide substantially improved statistics and an expanded $p_{T}$ range
over previously published results.

Jets are reconstructed at STAR\cite{star} with a midpoint cone algorithm\cite{Cone}
by clustering charged track and electromagnetic energy deposits within a cone (in azimuth, pseudorapidity) of radius 
$R = \sqrt{{\Delta \phi}^2 + {\Delta \eta}^2} = 0.4(0.7)$ in 2005(6).
Charged track momenta are measured by the Time Projection Chamber (TPC).
Neutral energy deposits are measured by the Barrel and Endcap Electro-Magnetic Calorimeters (BEMC, EEMC).
Beam Beam Counters (BBCs) are used in the range
$3.4 < |\eta| < 5$ on both sides of the collision point for monitoring of the beam luminosity.
A coincident signal between scintillator tiles from the East and West BBC
detectors forms STAR's baseline Minimum Bias (MB) trigger.

Events are triggered and analyzed if they included a High Tower (HT) or a Jet Patch (JP) above
set thresholds in addition to the MB condition.
The low(high) HT trigger in 2005 required each accepted event to 
have a tower in the BEMC ($\eta \times \phi = 0.05 \times 0.05$) with transverse energy $E_T \geq 2.6(3.5)$ GeV.  
The JP trigger in 2005 required $E_T \geq 4.5(6.5)$ GeV within a BEMC region 
$1\times1$ in $\eta$, $\phi$, and in 2006 required $E_T \geq 7.8(8.3)$ GeV
within the JP per accepted event.  Only JP triggered data is presented for 2006.

\begin{figure}
\begin{center}
\includegraphics*[width=6.65cm]{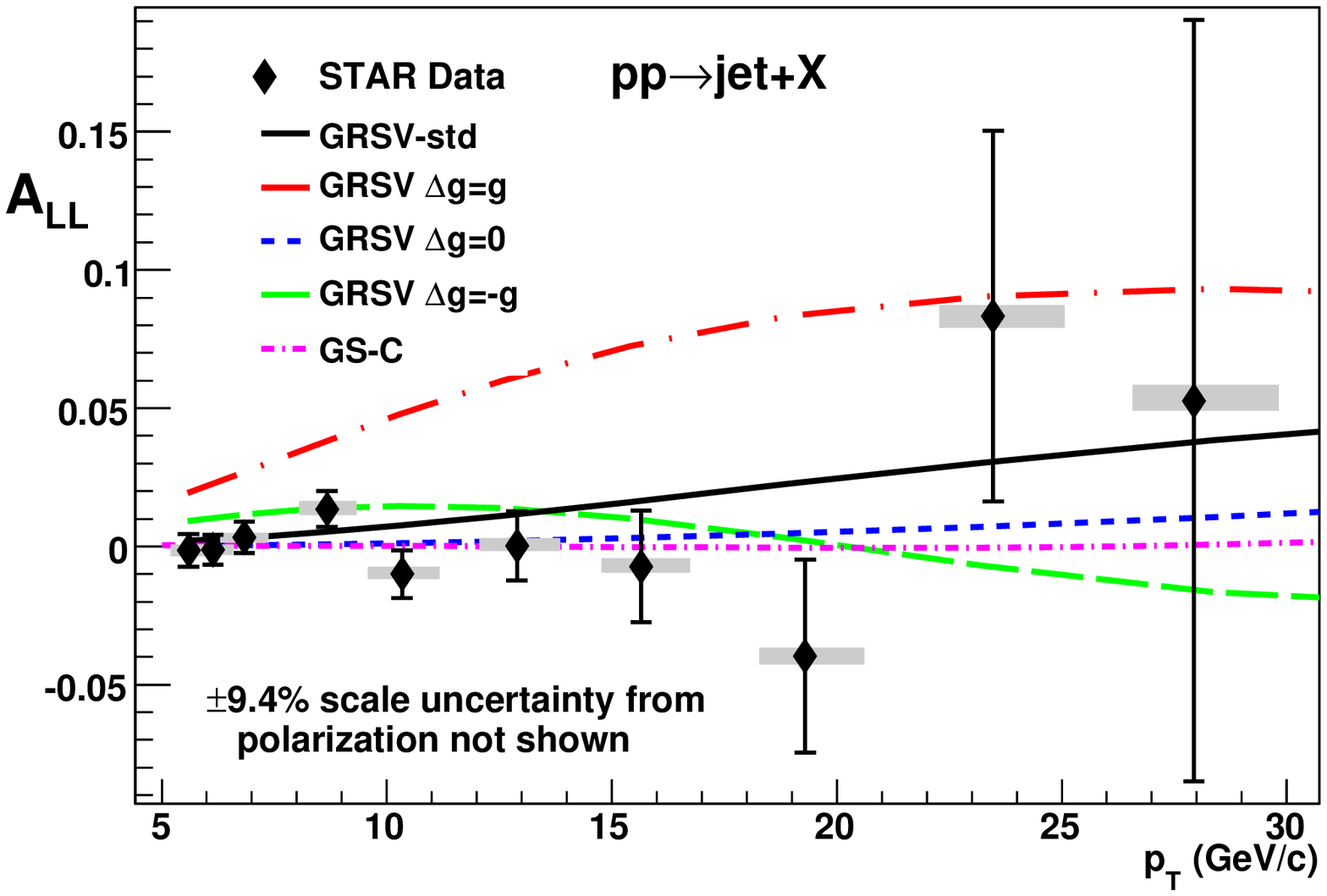}
\includegraphics*[width=6.65cm]{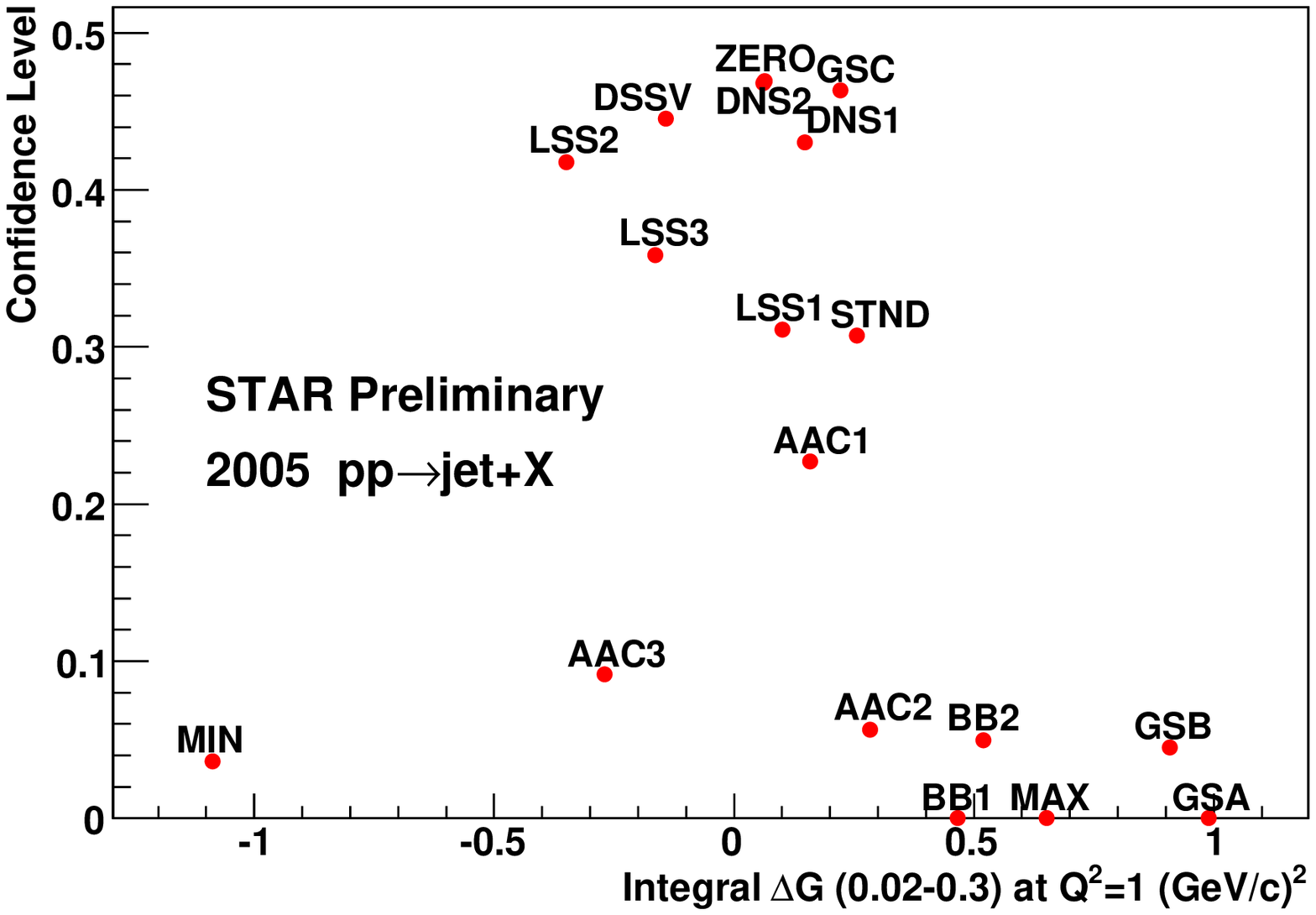}
\end{center}
\caption{Left: Final 2005 $A_{LL}$ as a function of corrected jet transverse momentum.
Error bars are statistical uncertainties only. Grey bands represent the
systematic uncertainties, excluding an overall uncertainty for
beam polarization.  Right: Confidence Level calculations for all available global fits of world
pDIS data.}
\label{2005ALL}
\end{figure}

\begin{figure}
\begin{center}
\includegraphics*[width=6.65cm]{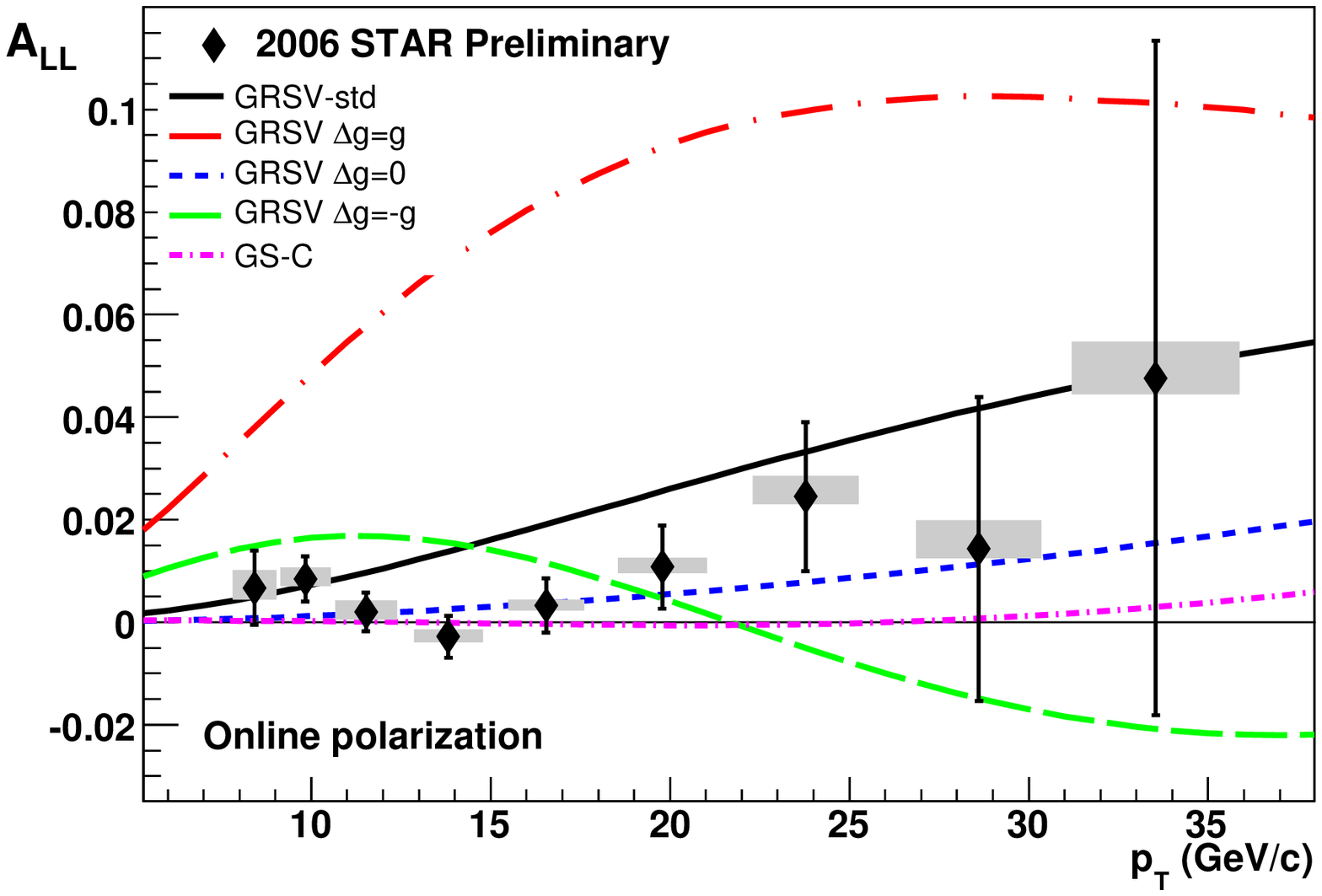}
\includegraphics*[width=6.0cm]{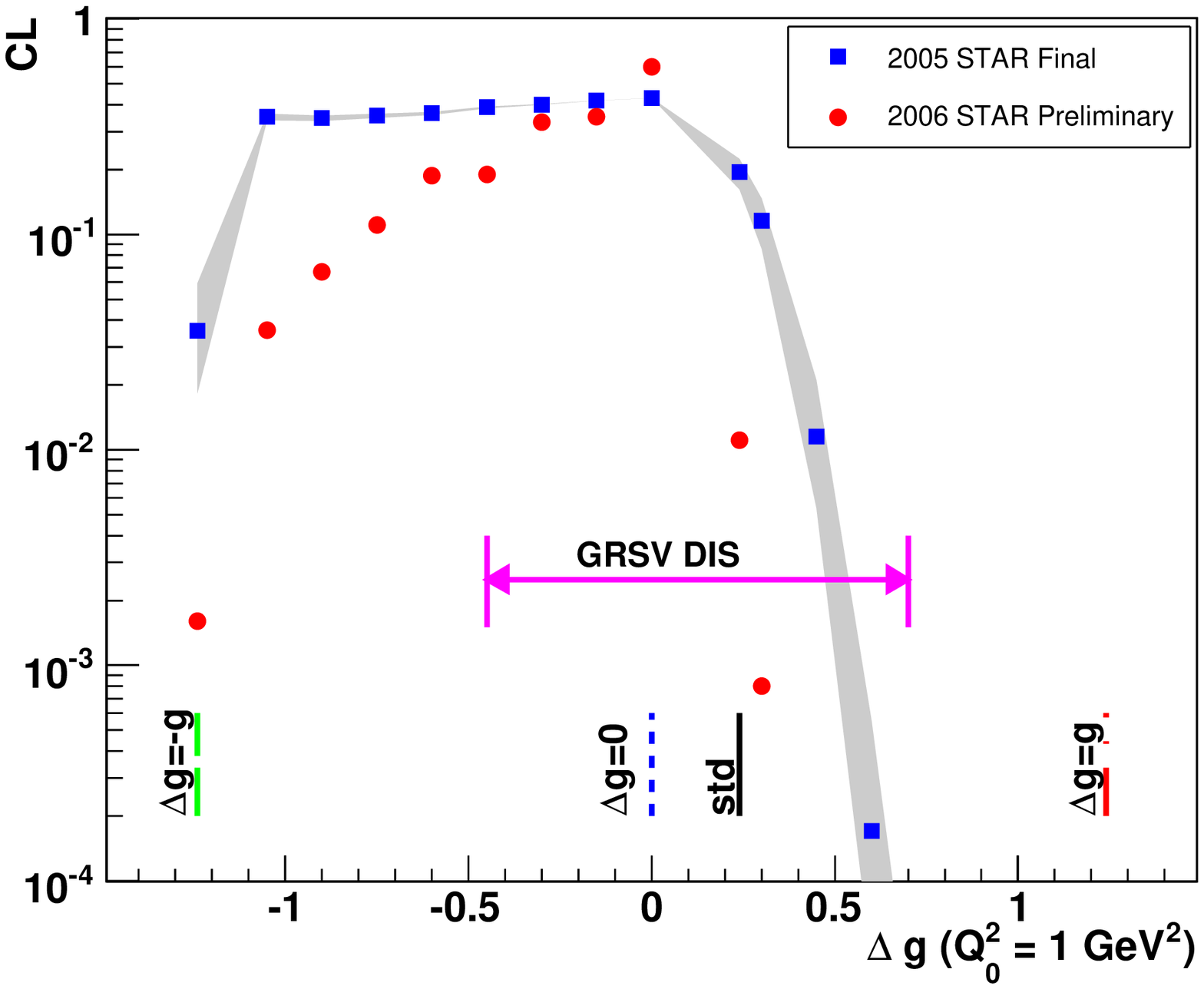}
\end{center}
\caption{Left: Preliminary 2006 $A_{LL}$ as a function of corrected jet transverse momentum.
Error bars are statistical uncertainties only. Grey bands represent the
systematic uncertainties, excluding an overall uncertainty for
beam polarization.  Right: Confidence Level calculations for GRSV\cite{grsv} $\Delta g$ models 
ranging from $\Delta g=-g$ to $\Delta g=g$.  The x-axis represents the integral of x from $0\rightarrow1$
of $\Delta g(x)$ for that model at the input scale $Q^2$=1.0 $GeV^2$.  Overlaid is the 1-$\sigma$
range around the best fit to global data, GRSV-STD.  The grey band is the $9.4\%$ scale uncertainty
from beam polarization.}
\label{2006ALL}
\end{figure}

Figure \ref{2005ALL}(\ref{2006ALL}) gives our measured $A_{LL}$ from 2.0(4.7) $pb^{-1}$ of data
taken during 2005(6).  
Typical beam polarization values during 2005(6) data-taking were 50(55-60)\%.
For both samples, the leading systematic uncertainty comes from the trigger and jet reconstruction biases. 
Differences between observed and true jet $p_T$ are estimated using PYTHIA\cite{pythia} and GEANT\cite{geant} simulations and 
result in corrections being applied to the jet $p_T$ values in Figs. \ref{2005ALL} and \ref{2006ALL}.
Additionally, the trigger conditions at STAR can artificially bias our sample towards
particular flavors of partonic collisions.  
A conservative systematic is evaluated to account for this effect which incorporates all the allowable 
$\Delta G$ models.  The method applied here is discussed in more detail in \cite{2005paper}.
Other less significant systematic uncertainties are also estimated, including relative luminosity
uncertainties, beam background event inclusion, and residual transverse beam components.

The $A_{LL}$ curves in Fig. \ref{2005ALL} are derived from NLO fits to world polarized DIS data 
by two separate theory groups, GRSV\cite{grsv} and GS\cite{gs}.  
Within the GRSV framework, the best fit to available pDIS is STD while $\Delta g$ =g, $\Delta g$ = 0,
and $\Delta g$ =-g are fits determined by varying the gluon polarization at the initial scale
$Q^2$ = 0.4 $GeV^2$.
GS-C is also shown and stands apart from the others in that it contains a node at $x\sim 0.1$
for $Q^2 = 4$ $GeV^2$ and has a small net contribution within the x-range of the measurement.
STAR's x-range, $0.03<x<0.3$, accounts for $\sim$ 50\% of the total $\Delta G$
integral for GRSV-STD.

Figure \ref{2006ALL} shows the confidence level (CL) for the comparison of the 2005/6 
data and theory for each of the GRSV parameterizations.  
Stratmann and Vogelsang\cite{percomm}  provided STAR with additional $\Delta G$
models lying between the 4 major GRSV models plotted on the left. 
In Fig. \ref{2005ALL}, CLs are calculated for GRSV and GS and all other available global analyses.
Models with large positive or negative gluon contributions within STAR's x-range are particularly constrained.

Preliminary 2006 and final 2005 inclusive jet $A_{LL}$ results from STAR have been presented.
The 2005(6) data represent 2(4.7) $pb^{-1}$ of data with typical beam polarizations around $50(\sim55-60)\%$.
A confidence level analysis has been shown which constrains the allowable theoretical 
gluon parameterizations.
A recent global analysis has been performed by the DSSV\cite{dssv} theory group that has included for 
the first time RHIC pp data with pDIS and pSIDIS data in their NLO fit.
The resulting fit constrains the gluon spin contribution to be small in the accessible experimental 
kinematic range, and contains a node near $x\sim0.1$ with a functional form opposite in sign to that of GS-C's.
STAR's preliminary 2006 jets measurement was used in their analysis and provides the leading constraint
on negative $\Delta g(x)$ over the range $0.05 < x < 0.2$, and comparable constraints to other measurements 
for positive $\Delta g(x)$ models.

\end{document}